\documentclass[conference]{IEEEtran}

\usepackage[brazil]{babel}
\usepackage[utf8]{inputenc}

\usepackage{cite}
\usepackage{amsmath,amssymb,amsfonts}
\usepackage{graphicx}
\usepackage{textcomp}

\usepackage{algorithm}
\usepackage{algpseudocode}

\floatname{algorithm}{Algoritmo}

\usepackage{comment}
\usepackage{url}

\def\BibTeX{
	{\rm B\kern-.05em{\sc i\kern-.025em b}\kern-.08em
    T\kern-.1667em\lower.7ex\hbox{E}\kern-.125emX}
}

\begin{document}

\title{
	O Problema do Roteamento de Interligações Elétricas em Circuitos Integrados
}

\author{
	\IEEEauthorblockN{Tiago Matos Santos}
	\IEEEauthorblockA{\textit{Instituto de Computação} \\
	\textit{Universidade Federal da Bahia}\\
	Salvador, Brasil \\
	tiagomatos@ufba.br}
}

\maketitle

\def\abstractname{Abstract}

\begin{abstract}
Integrated circuit design automation tools are essential for the feasibility of complex designs with millions of transistors. One of the steps performed within the process is the routing of interconnections between components of a circuit. This problem, which also aims to optimize the utilization of connection resources, has been shown to be NP-Complete and requires heuristic algorithms to look for the best achievable solutions. In this work, we present a definition of this problem in context with a brief review of existing solutions in the literature. Then, we propose a methodology for the development of an original algorithm, which aims to differentiate itself, in certain domains, from the solutions already proposed.
\end{abstract}

\def\abstractname{Resumo}

\begin{abstract}
Ferramentas de automação do projeto de circuitos integrados são fundamentais para a factibilidade de desenhos com complexidade da ordem de milhões de transístores. Uma das etapas executadas neste processo é a de roteamento das interconexões entre os componentes de um circuito.
Este problema, que objetiva também a otimização na utilização dos recursos de conexão, foi demonstrado ser NP-Completo e requer algoritmos heurísticos para a busca das melhores soluções alcançáveis. Neste trabalho, apresentamos a definição deste problema em contexto com uma breve revisão das soluções existentes na literatura. Em seguida, propomos uma metodologia para o desenvolvimento de um algoritmo original, que visa diferenciar-se das soluções já propostas, em determinados domínios.
\end{abstract}


\section{Introdução}
Circuitos integrados são a base para os modernos dispositivos eletrônicos e de computação que nos permeiam atualmente. Sua complexidade de projeto vem aumentando de forma exponencial desde sua introdução há mais de 50 anos \cite{wolf2008modern}. Atualmente, a tecnologia nos permite a fabricação de transistores com dimensões da ordem de 10 nm. Assim, uma pequena área de um circuito integrado pode conter milhões destes componentes \cite{wikichip10nm}.

Com a evolução da tecnologia de fabricação, houve também avanços nas ferramentas de projeto dos circuitos integrados, fundamentais para a factibilidade de desenhos com tamanha complexidade \cite{sherwani2014algorithms}. Uma nova sub-área de pesquisa surgiu, a de automação do projeto eletrônicos, do inglês electronic design automation (EDA). A ferramentas de EDA se inserem em diversas etapas existentes no desenvolvimento de equipamentos eletrônicos, em especial, são indispensáveis nas etapas de descrição e síntese digital de circuitos, no posicionamento e interconexão de componentes, e na geração das geometrias usadas na fabricação nos wafers de silício \cite{lavagno2017electronic}.

Neste trabalho, damos enfoque à etapa do roteamento de interligações entre os componentes presentes em um circuito integrado, que é caracterizada pela necessidade de obtenção resultados otimizados. Isso nos leva a uma classe de problemas que são NP-completos ou NP-difíceis a depender de sua formulação específica, assim sendo, tais problemas devem ser enfrentados de forma heurística \cite{maurer1990tutorial, sherwani2014algorithms}.

\section{Descrição Formal do Problema}
\label{sec:descricao}

Uma abordagem para solucionar problema é conhecida como roteamento por canal (channel routing), originalmente proposta em \cite{hashimoto1971channel}, que trabalha impondo restrições aos possíveis caminhos para as interconexões. Na prática, ao invés de explorar todo o espaço livre entre os componentes já posicionados em um circuito, esta abordagem cria os chamados canais nestes espaços, retângulos com terminais de interligação apenas em suas extremidades, que serão utilizados para a efetivação do roteamento.

\subsection{Algumas definições}

Aqui, são dadas algumas definições para termos relativos ao problema, que são de uso comum na literatura \cite{maurer1990tutorial}.

\begin{itemize}
	\item Células: são representações para os componentes do circuito que consistem de retângulos que podem estar posicionados em diversas posições de plano bidimensional.
	\item Terminais (ou pinos): Representam posições nas células (ou no canal) para os quais as interconexões devem ser estabelecidas. Podemos indicá-los por pontos nas bordas da célula (ou do canal).
	\item Net: Uma coleção de terminais interconectados por um mesmo caminho elétrico, a conexão também é parte da net.
	\item Interligação (ou conexão): Caminho conecta em um mesmo potencial elétrico terminais, vias e segmentos.
	\item Segmentos: Partes de uma interligação que são segmentos de retas finitas.
	\item Camadas: Acrescentam uma dimensão a mais, na direção de baixo para cima, ao roteamento. Terminais geralmente ficam na primeira camada e as demais interligações podem estar dispostas em diversas camadas.
	\item Vias: Conexões de segmentos entre diferentes camadas.
	\item Canal: Uma área retangular destinada à disposição das interconexões elétricas. Contém um conjunto fixo de terminais apenas em uma orientação (horizontal por exemplo). Conectando-se terminais existentes nas células aos de um canal, será possível estabelecer as \textit{nets} necessárias para interligação de células.
\end{itemize}

\begin{figure}[htbp]
	\centerline{
		\includegraphics[width=0.45\textwidth]{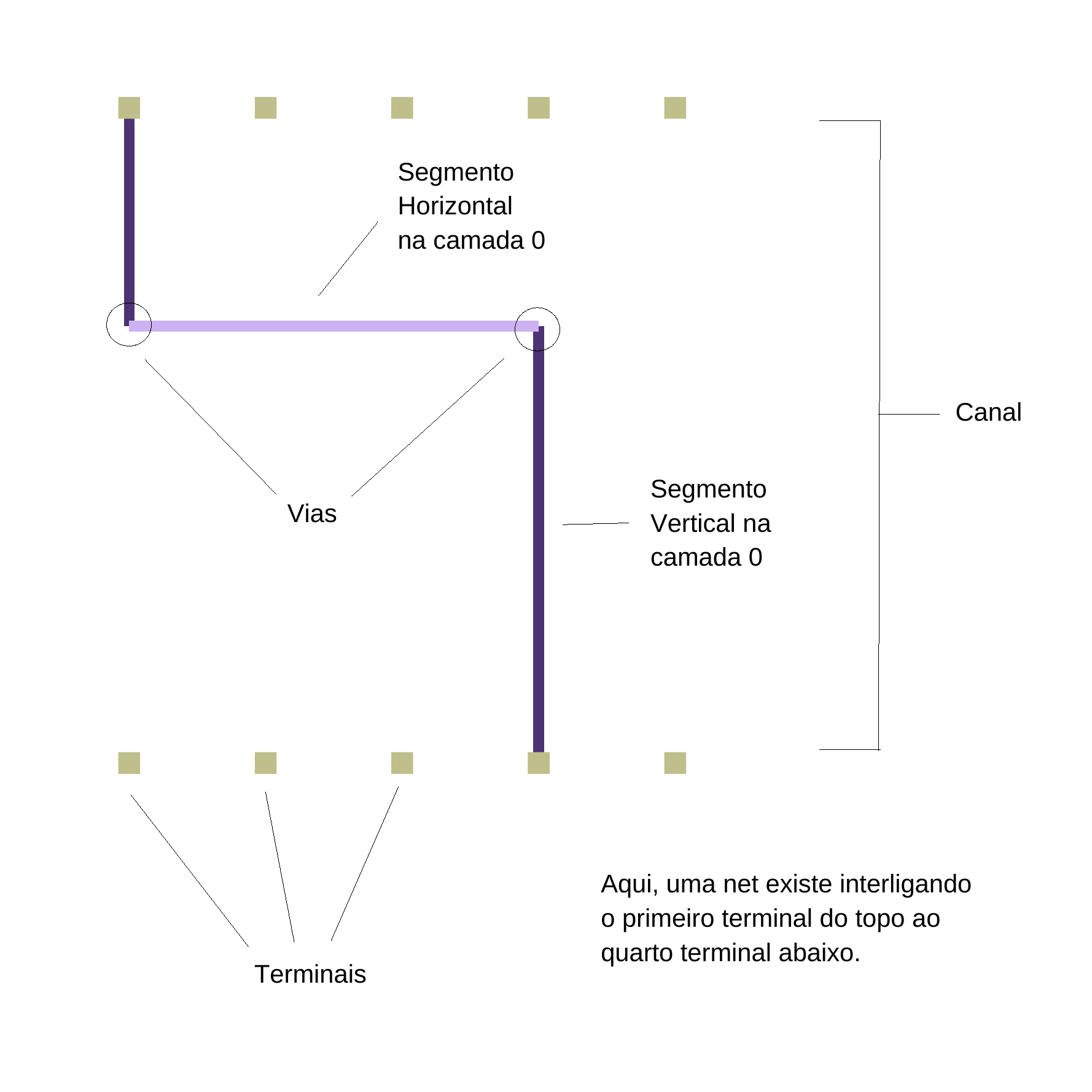}
	}
	\caption{Indicações dos elementos resultantes no processo de roteamento por canal de interconexões.}
	\label{fig:rotas-legendas}
\end{figure}

Na Figura \ref{fig:rotas-legendas}, algumas ilustrações são dadas para os itens descritos anteriormente, com suas devidas indicações.

\subsection{Especificação}

Tomamos como entradas para um algoritmo de roteamento um arranjo de terminais em duas dimensões, uma topologia de nets a serem roteadas e, opcionalmente, algumas restrições para possíveis caminhos de interconexões elétricas. O algoritmo deve então estabelecer conexões para todas as nets do circuito, de modo que nets diferentes não entrem em contato, enquanto minimiza as camadas necessárias e os comprimentos das conexões.

\subsection{Simplificações}

Em nossa especificação, devido ao escopo limitado deste trabalho, vamos manipular o problema adotando simplificações, algumas delas também adotadas em publicações relacionadas.

\begin{itemize}
	\item Todas as conexões são de formatos retilíneos, permitindo-se ao longo da rota apenas varições de 90º na orientação.
	\item Camadas apresentam, alternadamente, canais em apenas uma orientação, vertical ou horizontal.
	\item A área do canal é dividida por uma grade em regiões unitárias, que serão preenchidas pelos objetos.
	\item As partes horizontais ou verticais das conexões seguem as linhas do grid.
	\item Terminais ocupam pontos do grid.
\end{itemize}

\section{Justificativa para a Classificação do Problema como NP-Completo}

O fato de que o problema de roteamento baseado em canais é NP-Completo foi demonstrado em \cite{szymanski1985np}. Onde foram utilizadas, além das simplificações do problema citadas na seção~\ref{sec:descricao}, uma restrição a apenas duas camadas no circuito físico.
No trabalho citado, a construção utilizada foi uma redução do problema da 3-satisfatibilidade, que sabe-se de antemão ser NP-Completo de acordo com \cite{Karp1972}, para o problema em questão.

\section{Breve revisão da literatura}

O problema do roteamento por canal tem recebido diversas visitas da comunidade acadêmica ao longo dos anos. E as soluções para ele  tem sido a base para o projeto de muitas ferramentas de EDA que chegaram a ser disponibilizadas \cite{maurer1990tutorial}. A seguir, são enumerados os algoritmos de maior destaque na literatura.

\subsection{Algoritmo ''left edge''}

O trabalho original que propôs o roteamento por canal apresentou o primeiro algoritmo \cite{hashimoto1971channel}. Voltado a apenas duas camadas, numa época em que sua aplicação estaria restrita apenas às placas de circuito impresso.

Partindo dos espaços livres entre componentes numa placa que se estendem de uma borda a outra, que são a primeira coisa a ser atribuída neste algoritmo, uma atribuição de canais é definida para então ser roteado de uma maneira bastante direta e simples no que viria a ser conhecido como algoritmo ''left edge''. Nele, segmentos horizontais de circuito são processados, em ordem, da parte de cima até a de baixo do canal, onde a cada estágio as nets são atribuídas aos segmentos disponíveis.

\subsection{Algoritmo ''dogleg''}

Este algoritmo apresentou um bom comportamento em utilizações práticas. Ele faz uso do chamado ''dogleg', que é um segmento de fio vertical que conecta dois outros segmentos horizontais. Desta forma, permite-se que uma net possa utilizar dois ou mais segmentos horizontais, através do uso de via extras. Isso permite que o circuito obtido fique mais compacto \cite{Deutsch1976dogleg}.

\subsection{Algoritmo guloso}

Apresentado como rápido, simples e efetivo. Usa as restrições de alternância na orientação dos canais na camada e da divisão em grade do canal como apresentadas na seção~\ref{sec:descricao}. Este algoritmo posiciona segmentos de fio no canal da esquerda para a direita, coluna a coluna até que esta se complete, enquanto tentar maximizar a utilidade dos segmentos introduzidos, de acordo com heurísticas gulosas \cite{Rivest1982greedy}.

\section{Metodologia}

\begin{figure}[htbp]
	\centerline{
		\includegraphics[width=0.45\textwidth]{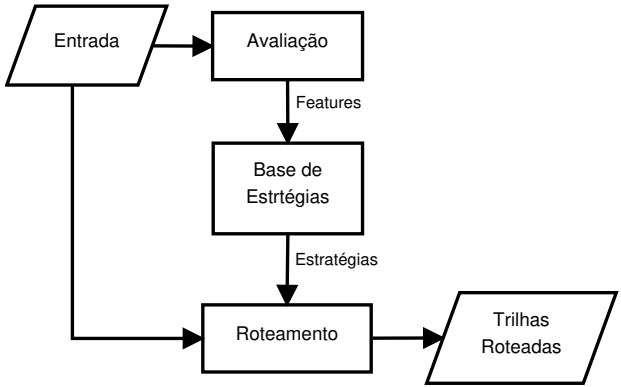}
	}
	\caption{Diagrama da solução, onde busca-se melhores estratégias de roteamento de acordo com o tipo da entrada inserida.}
	\label{fig:diagrama-geral}
\end{figure}

Nossa abordagem para propor uma solução original para o problema vai na mesma linha dos algoritmos heurísticos. Como diferencial, propomos um direcionamento no espaço de busca das soluções, que será guiado por meio de características extraídas das entradas específicas. O relacionamento entre essas características e direção a ser adotada será dado a partir de comparações com um banco de dados pré-computado que estabelece algumas probabilidades de sucesso dadas determinadas combinações de características e direções. A Figura \ref{fig:diagrama-geral} resume a ideia a ser implementada.

Em outras palavras, dada a especificidade de cada entrada, o algoritmo escolherá ou modificará uma determinada estratégia de posicionamento das trilhas no espaço do canal de roteamento. Com a esperança de alcançar resultados melhores que heurísticas estáticas.

\section{Desenvolvimento}

\subsection{Programa}

\begin{figure}[htbp]
	\centerline{
		\includegraphics[width=0.45\textwidth]{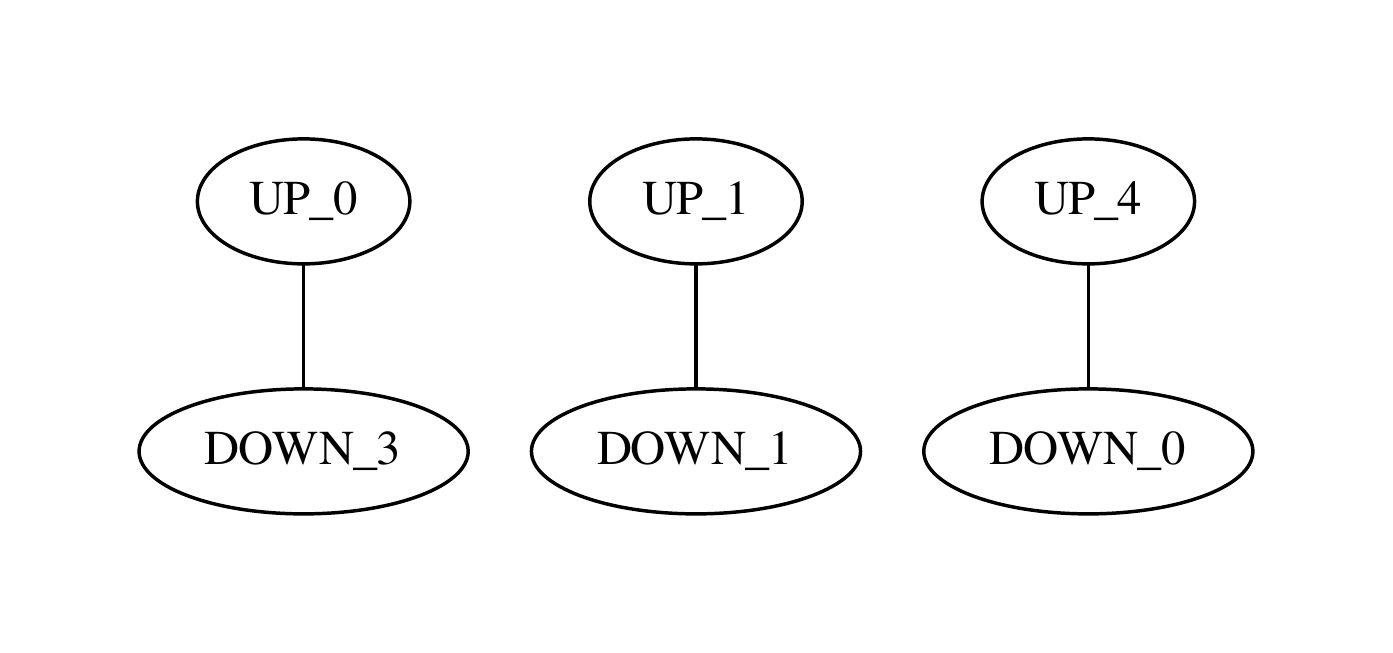}
	}
	\caption{Grafo de uma netlist. Os terminais na parte superior do canal são indicados com o prefixo UP, os da parte inferior possuem o prefixo DOWN. Arestas indicam a conectividade à mesma \textit{net}.}
	\label{fig:grafo-netlist-exemplo}
\end{figure}

Optou-se por implementar um programa completo para a solução de instâncias do problema. O problema é capaz de ler as netlists a partir de um arquivo textual em formato próprio, e a carrega para as respectivas  estruturas de dados em memória, que passam então por várias etapas de processamento. Uma netlist pode ser representada como um grafo não dirigido, em que cada vértice representa um terminal, e cada aresta indica que existe uma conexão entre os respectivos terminais, isto é, eles fazem parte da mesma net. Um exemplo disto é o grafo da Figura \ref{fig:grafo-netlist-exemplo}.

\begin{figure}[htbp]
	\centerline{
		\includegraphics[width=0.375\textwidth]{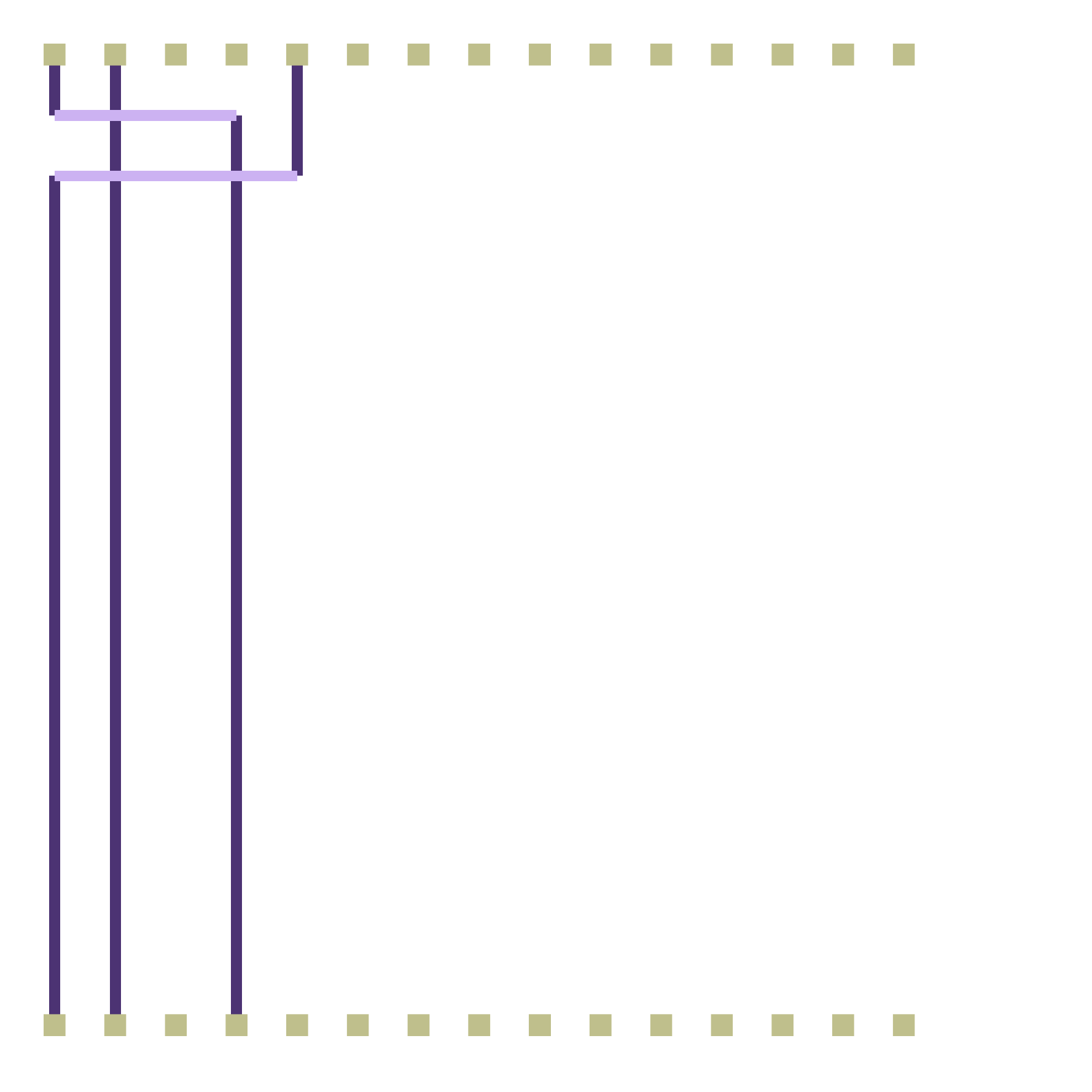}
	}
	\caption{Exemplo de um roteamento em canal para 3 nets.}
	\label{fig:rotas-exemplo}
\end{figure}

O programa foi escrito na linguagem Python 2.7 e é capaz de exportar grafos no formato dot da biblioteca Graphviz, que são úteis para visualização dos mesmos. Além disso, ele gera como saída arquivos no formato PDF contendo uma representação gráfica das trilhas de roteamento geradas para o canal. Um exemplo está na Figura \ref{fig:rotas-exemplo} que é o roteamento para a topologia da Figura \ref{fig:grafo-netlist-exemplo}. Isto foi possível com a utilização da biblioteca gráfica Cairo.

Para sua execução, basta invocá-lo em um terminal de texto com o comando: \texttt{python2 rotear.py <circuito-exemplo.netlist>}. No diretório com os códigos fonte distribuídos há alguns exemplos de circuitos no formato \texttt{'.netlist'}.

\subsection{Técnicas Utilizadas}

\begin{figure}[htbp]
	\centerline{
		\includegraphics[width=0.35\textwidth]{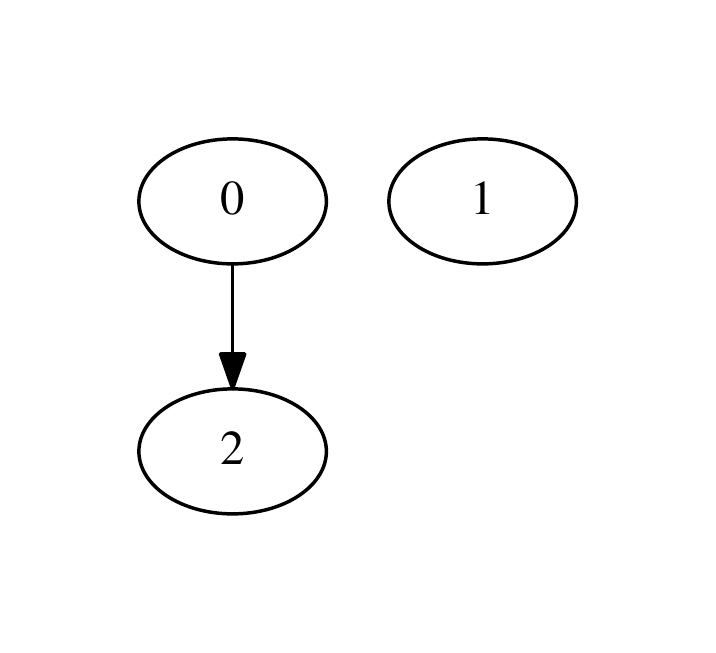}
	}
	\caption{Grafo de restrições verticais para uma netlist, onde cada vértice representa uma net. No exemplo, a net relativa ao vértice 0 restringe de alguma forma o posicionamento de segmentos do vértice 2.}
	\label{fig:grafo-vcg-exemplo}
\end{figure}

Restrições para o posicionamento de segmentos em determinadas posições do grid pode sem derivadas a partir do cálculo dos grafos de restrições horizontais (HCGs) e grafos de restrições verticais (VCGs) \cite{kahng2011vlsi}. Tais grafos servem como guia para a criação e o posicionamento de segmentos das trilhas do circuito. Um exemplo está na Figura \ref{fig:grafo-vcg-exemplo}, que é o VCG para a netlist da Figura \ref{fig:grafo-netlist-exemplo}.

Além disso, propomos o uso de informações em nível macro da netlist de entrada para inferir, através de estratégias pré-computadas, co-orientações para a execução de nosso algoritmo. Neste trabalho, dado o caráter preliminar, optamos por utilizar a informação do particionamento dos terminas entre a parte esquerda e direita com relação ao meio do canal de roteamento. Esses valores são comparados com listas que retornam dicas para o posicionamento dos próximos segmentos de cadas trilha.

\subsection{Algoritmo}

O Algoritmo \ref{alg-roteamento} é implementado como a principal rotina do programa, e tem a responsabilidade de receber uma estrutura de dados para uma lista de nets como entrada, já carregada por outras sub-rotinas do programa, retornando ao final uma lista de trilhas (tracklist), que nada mais é que um conjunto de listas ordenadas de segmentos. A ordem estabelecida, no caso destes segmentos, é a de um percurso possível para a corrente elétrica. Temos aqui a função NOVA-LINHA no algoritmo, que serve para escolher uma nova linha para posicionamento dos segmentos horizontais que, não entre em conflito com outros segmentos horizontais já posicionados, e que busque um melhor aproveitamento do espaço do canal.

\begin{algorithm}
	\caption{Gerar o roteamento para uma netlist.}
	\label{alg-roteamento}
	\begin{algorithmic}[1]
		\Statex
		\Require NL (lista de conjuntos de terminais conectados).
		\Ensure TL (lista de conjuntos de segmentos de reta respectivo a cada conjunto da lista NL de entrada).
		\Statex
		\Procedure{Rotear}{NL, TL, sucesso}
			\State MINLINHA = 1
			\State MAXLINHA = máx. núm. de linhas de grid
			\State MINCOLUNA = 0
			\State MAXCOLUNA = máx. núm. de colunas de grid
			\Statex
			\State vcg = obterVcg(NL)
			\State hcg = obterHcg(NL)
			\State info = obterInfo(NL)
			\Statex
			\ForAll{net in NL}
				\State esq(net) = terminal mais à esquerda na net
			\EndFor
			\Statex
			\State \Call{ORDENAR}{NL, esq(net)} \Comment{odenar NL por ordem de terminal mais à esquerda em cada conjunto}
			\Statex
			\State linha = MAXLINHA / 2
			\State coluna = 0
			\State camada = 0
			\State trilha = []
			\Statex
			\ForAll{net in NL}
				\State x0 = net.terminal[0].coluna
				\State y0 = net.terminal[0].linha
				\State xN = net.terminal[1].coluna
				\State yN = net.terminal[1].linha
				\Statex
				\If {terminais na mesma coluna}
					\State segmento = [camada, (x0, y0), (xN, yN) ]
					\State \Call{Anexa}{trilha, segmento}
				\Else
					\State segmento = [ (x0, y0), (x0, linha) ]
					\State \Call{Anexa}{trilha, camada, segmento}
					\State segmento = [ (x0, linha), (xN, linha) ]
					\State \Call{Anexa}{trilha, camada+1, segmento}
					\State segmento = [ (xN, linha), (xN, yN) ]
					\State \Call{Anexa}{trilha, camada, segmento}
					\State linha = NOVA-LINHA(vgc, hcg, info, TL)
					\If {linha == NIL}
						\State sucesso = False
						\State \Return
					\EndIf
				\EndIf
			\EndFor
			\Statex
			\State sucesso = True
			\State NL = trilha
			\Statex
		\EndProcedure
		
	\end{algorithmic}
\end{algorithm}

\section{Resultados}

As principais métricas para verificação da optimalidade das soluções encontradas são o menor número de canais necessários, seguido pela menor quantidade de linhas necessárias no canal como um todo, e em último lugar os menores comprimentos totais de segmentos. A versão implementada até o momento está restrita a dois canais. Quanto à racionalização do uso de linhas, ela se iguala ou tem resultados inferiores a exemplos da literatura. Acreditamos que a configuração do banco de estratégias, etapa ainda não plenamente concluída, possa nos levar a inúmeras melhorias, e isto é algo a se objetivar em um trabalho futuro.

\section{Conclusão}

Neste relatório, apresentamos um problema relevante e de grande interessante acadêmico e industrial, pertencente à classe NP-Completo, que é o roteamento por canal de interconexões em circuitos integrados. Fizemos uma descrição do problema, com suas simplificações pertinentes, e introduzimos a literatura referente à sua análise de intratabilidade. Além disso, após uma breve revisão de literatura, onde observamos as principais abordagens pesquisadas e implementadas ao longo do anos. Propusemos um algoritmo original, que tem por objetivo superar o desempenho das soluções consagradas, ao menos em um conjunto de interesse de instâncias, ao inserir randomizações nas estratégias de roteamento relativas a características peculiares de cada entrada.

\nocite{*}

\bibliographystyle{IEEEtran}
\bibliography{IEEEabrv,referencias}

\end{document}